# Molecular nitrogen acceptors in ZnO nanowires induced by nitrogen plasma annealing

C. Ton-That[*], L. Zhu, M. N. Lockrey and M. R. Phillips

*School of Mathematical and Physical Sciences, University of Technology Sydney, PO Box 123, Broadway, NSW 2007, Australia*

B. C. C. Cowie, A. Tadich and L. Thomsen

*Australian Synchrotron, 800 Blackburn Road, Clayton, VIC 3168, Australia*

S. Khachadorian, S. Schlichting, N. Jankowski and A. Hoffmann

*Institut für Festkörperphysik, Technische Universität Berlin, Hardenbergstr. 36 10623 Berlin, Germany*

*Corresponding author. Email: cuong.ton-that@uts.edu.au

## Abstract

X-ray absorption near-edge spectroscopy (XANES), photoluminescence, cathodoluminescence and Raman spectroscopy have been used to investigate the chemical states of nitrogen dopants in ZnO nanowires. It is found that nitrogen exists in multiple states: $N_O$, $N_{Zn}$ and loosely bound $N_2$ molecule. The work establishes a direct link between a donor-acceptor pair (DAP) emission at 3.232 eV and the concentration of loosely bound $N_2$. These results confirm that $N_2$ at Zn site is a potential candidate for producing a shallow acceptor state in N-doped ZnO as theoretically predicted by Lambrecht and Boonchun [Phys. Rev. B 87, 195207 (2013)]. Additionally, shallow acceptor states arising from $N_O$ complexes have been ruled out in this study.

**I. INTRODUCTION**

Nitride based wide-bandgap semiconductors have facilitated and expedited major advances in high-power and high-frequency electronics and full solar spectrum photovoltaic cells. This progress was achieved following the development of specialized fabrication techniques that have enabled the routine growth of *p*-type nitride semiconductors with different doping levels, which for many years appeared to be an insurmountable challenge due to self-compensation by intrinsic point defects [1]. Oxide based semiconductors have recently emerged as a very promising alternative to nitride semiconductors and have the capacity to significantly enhance the performance and applications of wide-bandgap semiconductor devices. In particular, ZnO with unique optoelectronic properties and several fundamental advantages over nitrides, has opened the door to a wide range of possible applications [2]. Like its nitride counterparts, the benefits of ZnO can only be realized once a reliable *p*-dopant and associated fabrication methods are established.

Among acceptor candidates for ZnO, nitrogen is considered to be the natural choice; however, to date it has not produced *p*-conductivity either because of the low nitrogen solubility or the formation of compensating defects [3]. Nevertheless, a shallow acceptor level has been reported in N-doped ZnO bulk and nanostructures [4, 5]. On the basis of a donor-acceptor pair (DAP) recombination at ~ 3.24 eV, the binding energy of the acceptor level was determined to be 170 ± 40 meV [4, 6, 7]. Recent extensive theoretical work, however, shows that N substituting O ($N_O$) leads to a deep acceptor level of 1.3 eV [8] or 1.6 eV above the valence band [9]. These discrepancies between the reported experimental and theoretical values raise questions about the exact chemical nature of the acceptor in N-doped ZnO. In this paper, we report the experimental observation of $N_2$ formation in ZnO nanowires by means of synchrotron-based X-ray Absorption Near-Edge Structure (XANES) and establish a direct link with an optical emission attributed to a nitrogen acceptor.

Several chemical states of N-related defects are electronically active in ZnO, acting as both donors and acceptors. Isolated nitrogen at O sites was initially detected by electron spin resonance (ESR) and attributed to the shallow acceptor in ZnO [5], whereas it has been established that $N_2$ on an oxygen site $(N_2)_O$ acts as a shallow donor, which compensates the *p*-type doping [10, 11]. In reality, the nature of the N-related acceptor state is complex and has until now remained unclear experimentally and controversial theoretically. Based on the combination of density functional calculations and absorption energies of N atoms on ZnO surfaces, Liu *et al.* [12] proposed the defect complex $N_O$-$V_{Zn}$ being a shallow acceptor. This complex is preferentially formed on the Zn-terminated surface during ZnO growth but the transition of $N_{Zn}$-$V_O$ to $N_O$-$V_{Zn}$ must overcome a large energy barrier of 1.1 eV. More recently, it has been predicted that $N_2$ molecule could be accommodated at a Zn site [13]. By comparing the first-principles calculation results and the ESR spectra of N-doped ZnO, Lambrecht and Boomchun [13] concluded that $(N_2)_{Zn}$ is the shallow double acceptor responsible for the DAP transition. To date, no experimental study has examined the interrelationships between the chemical states of nitrogen dopants and the optical signatures of shallow acceptors. Here we elucidate the opto-electronic properties of ZnO nanowires doped with specific nitrogen species identified using XANES spectroscopy. Nanowires were specifically chosen for this study as these low dimensional structures enable control over the formation of native defects during growth as well as uniform doping of nitrogen throughout their entire volume.

**II. EXPERIMENTAL DETAILS**

ZnO nanowires were grown by chemical vapour deposition in a gas-controlled furnace with mixed ZnO and graphite powder as source material [14]. Using this synthesis technique, the nanowire morphology and their defect chemistry could be controlled by varying the growth temperature (700°C – 900°C) in combination with the flow rate of oxygen carrier gas. To enhance nitrogen incorporation by vacancy doping [10], nanowires were grown under Zn-rich conditions, which lead to an intense defect emission band peaked peaking at 2.47 eV, attributable to $V_O$ [15]. Nitrogen doping was achieved by annealing nanowires at 300°C in nitrogen plasma with 230 V cathode bias. The nanowires were characterized using a FEI Quanta 200 Scanning Electron Microscope (SEM) with an attached cathodoluminescence (CL) spectrometer. High resolution photoluminescence (PL) spectroscopy was performed using the fourth harmonic generation (266 nm) of a Nd:YAG laser, the emitted light was dispersed by a monochromator with a 1200 lines/mm grating (spectral resolution 0.2 meV). XANES implemented in the Total Fluorescence Yield (TFY) mode was performed on the Soft X-ray Spectroscopy beamline at the Australian Synchrotron. The incident X-ray beam was parallel to the *c*-axis of the nanowires and linearly polarized. The photon energy scale was calibrated against the Au $4f_{7/2}$ peak at 84 eV from a clean gold film in electrical contact with the sample. Raman data were recorded using a LabRAM HR800 spectrometer (Horiba Jobin Yvon) with 458 nm laser excitation. The Raman data were collected in backscattering geometry with a spectral resolution better than 0.3 cm$^{-1}$. The laser power on the sample was tuned to 1 mW.

**III. RESULTS AND DISCUSSION**

Fig. 1 shows representative nitrogen *K*-edge XANES spectra of the as-grown and N-doped ZnO nanowires (an SEM image of the nanowires shown on the inset). The nanowires have a diameter ranging from 80 to 180 nm and a typical length of 0.6 μm. X-ray diffraction analysis (not shown) confirmed the nanowires as single-crystal nanostructures, aligned along [001] axis orientation. No changes in the nanowire morphology were detected after plasma annealing. The XANES spectra of N-doped nanowires could be deconvoluted into three Lorentzian peaks to represent resonant electron transitions from the N 1$s$ initial state to the final unoccupied N-related state of $p$-symmetry [11], after subtracting an arctangent step function that represents the transition of ejected photoelectrons to the continuum [16]. Examples of the peak deconvolution are shown in Figs 1(a, b). The components centred at 400.0, 400.7 and 404.5 eV (labelled $P_1$, $P_2$ and $P_3$, respectively) correspond to three chemical states of nitrogen. The strong resonant peak $P_1$ at 400.0 eV is at the same position as the resonance previously observed in N-doped bulk ZnO crystals [17] and has been assigned to the characteristic 1$s$ to 2$p\pi^*$ transition in the hybridized N-Zn orbital [18]. The $P_2$ peak at 400.7 eV is associated with the N 1$s$ to $\pi^*$ transition in the N-N species [19], which is a signature of molecular nitrogen. The energy and width of this resonance peak are very similar to the vibration structure of trapped $N_2$ previously observed in oxides and nitrides, such as ZnO [17], GaN [20] and InN [21], implanted with N ions. As a fingerprint technique for the bonding arrangement of a dopant in the lattice, the subtle variation of the $P_2$ parameters among the compounds confirms that this $N_2$ species interacts weakly with the surrounding matrix. The energy of the broad $P_3$ peak is similar to that of the characteristic absorption peak of $N_2O$ and has been attributed to the N 1$s$ to 2$p\sigma^*$ transition in N-O species [18]. Accordingly, $P_3$ is ascribed to N at Zn sites. The $P_2$ component is enhanced significantly with increasing plasma time, indicating more nitrogen is present as molecular species. Further evidence for the formation of bound $N_2$ in N-doped nanowires is the emergence of high-energy Raman lines at 2269 and 2282 cm$^{-1}$ after plasma annealing, as shown in Fig. 1(d). These Raman peaks, close to the vibrational mode of free (gaseous) $N_2$ at 2329 cm$^{-1}$, have been attributed to lattice-bound $N_2$ [22, 23]. The Raman spectra of the nanowires in the low-energy range can be found in Appendix A. Both undoped and N-doped nanowires show typical ZnO lattice modes. No broadening of the modes is observed in N-doped nanowires, indicating the structural quality is not reduced by the incorporation of nitrogen.

Fig. 2(a) shows the CL spectra (5 kV, 0.52 nA) of individual ZnO nanowires, recorded from the same samples as in Fig. 1. The CL spectrum of the as-grown nanowires consists of a $D^0X$ emission at 3.355 eV and phonon replicas of free excitons (FX–LO and FX–2LO) at 3.313 and 3.237 eV, respectively. The absence of the zero phonon FX emission is due to efficient photon self-absorption, while the transmission of the sub-bandgap FX–$n$LO signals is relatively unaffected. Nitrogen doping produces two significant changes to the CL spectra. First, a 4 meV redshift of $D^0X$ emission at 3.355 eV is observed; this peak at 3.351 eV is ascribed to neutral acceptor bound exciton ($A^0X$) in N-doped ZnO. Second, a new peak at 3.232 eV emerges that is highly overlapped with the FX-2LO, leading to the pronounced decrease in the $I_{FX\text{-}LO} / I_{FX\text{-}2LO}$ ratio after N-doping which should be constant at a given temperature. Additionally, the 3.232 eV peak has a Gaussian-like shape following N incorporation, while FX-$n$LO phonon replicas are highly asymmetric peaks due to the Maxwellian distribution of the free exciton kinetic energies involved in the phonon scattering [24]. Accordingly, this emission is attributed to DAP transition. It is noteworthy that annealing the nanowires in Ar gas does not alter the CL properties of the nanowires. Further support for this assignment comes from the temperature dependence of FX replica and DAP peak positions (see Appendix B for the temperature-dependent characteristics of DAP and bound exciton peaks). Numerical simulation shows the variations of FX-LO and FX-2LO peak positions follow closely the standard equations (presented in Fig. 6) with a fitting parameter $E_{FX}$ = 3.378 eV (for bulk ZnO, $E_{FX}$ = 3.375 eV [25]). Due to the thermal energy $½k_BT$ contribution to $E_{FX\text{-}2LO}$ [26], the FX-2LO and N-related DAP occur essentially at the same energy above 140 K, but gradually separate with decreasing temperature. The DAP transition in N-doped nanowires exhibits a slight blue shift with increasing excitation density (Fig. 7). In this case a weak blue shift is not unexpected since only a small number of near donor-acceptor pairs exist due to the low concentration of nitrogen acceptors.

To further investigate the signatures of nitrogen acceptor in nanowires, high resolution PL spectroscopy at 7 K was conducted on the nanowire ensembles [Fig. 2(b)]. The emergence of a clear and resolvable peak at 3.236 eV following N-doping further confirms the assignment of this emission to a N-related DAP transition. As expected, the PL spectra of both undoped and N-doped nanowires at 7 K are dominated by bound exciton emissions and their phonon replicas. Moreover – and importantly– the bound exciton emission of ZnO nanowires exhibits significantly stronger coupling to optical phonons after N-doping, as manifested by the strong 2LO relative to the 1LO peak. This characteristic is typical of an acceptor-related transition [6], and confirms that nitrogen is the dominant acceptor in N-doped nanowires.

Now, we turn to the identity of the acceptor in the DAP transition. In the first report of ZnO-based LED, Tsukazaki *et al.* [27] reported that a nitrogen concentration of $10^{20}$ cm$^{-3}$ produced only $10^{16}$ holes/cm$^3$ in *p*-type ZnO films grown by MBE. In Fig. 3, we plot the integrated intensities of the $P_1$, $P_2$ and $P_3$ components together with the CL I(3.232 eV)/I(LO) peak intensity ratio as a function of plasma time. The difficulty in separating the overlap contributions of DAP and 2LO can be overcome by using the integrated intensity of the 3.232 eV peak; the increase in this peak intensity signifies the intensity of the N-related DAP. The XANES intensities represent the numbers of $N_O$, loosely bound $N_2$ and $N_{Zn}$ in the nanowires. Among the three chemical states of nitrogen in the nanowires, the $P_2$ center is the only candidate that can be responsible for the increase in the DAP intensity. The trend of $P_1$ has no association with the DAP emission, while $P_3$ can be ruled out since the substitution of N atom, which contains three more valence electrons than Zn, would cause $N_{Zn}$ or its complex to act as a donor [12]. The increase in the I(3.232 eV)/I(LO) ratio is not exactly proportional to the change of $P_2$ intensity since the intensities of the signature CL peaks are dependent on the capture cross section and concentration of all radiative and non-radiative recombination centers as well as the excitation conditions due to saturation effects [15]. It is noteworthy that the $P_3$ donor increases slightly in concentration with plasma time and could act as a compensation centre in N-doped nanowires. The enhancement of the DAP intensity is not associated with the overall nitrogen content, which remains almost unchanged for anneals longer than 30 mins. The direct correlation between the $P_2$ and I(3.232 eV) strongly indicates the involvement of loosely bound $N_2$ in the DAP transition and supports the $N_2$ acceptor model [13, 28], in which $N_2$ at a Zn site is weakly bound to the lattice and retains most of its molecular characteristics. Atomic N at the O site or its complexes have been theoretically predicted to be the shallow acceptor [10, 12]; however, this possibility can be ruled out for ZnO nanowires. While interstitial $N_2$ is known to readily form in GaN [29], the presence of such species in a large fraction is unexpected for ZnO and this could provide a new route for enhanced nitrogen acceptor incorporation. It is significant that data recorded in the surface sensitive total electron yield (TEY) mode exhibit completely different XANES spectra (not shown) with abnormally high backgrounds, presumably due to nitrogen-containing contamination on the nanowire surface. The use of the TFY detection mode as a bulk probe guarantees that the detected $N_2$ is representative of the bulk property of the nanowires. In addition, the self-absorption effects of nitrogen on the TFY data could be avoided because of small diameter of the nanowires.

Under the nitrogen plasma conditions used, ionized atomic and molecular nitrogen species generated in the plasma are accelerated towards the samples. The probability of a $N_2^+$ disassociating into 2 N atoms depends on the incident energy when the ion strikes a solid surface [30]. The energy required to break the bond between two N atoms in a $N_2^+$ molecule is about 9 eV [31], thus the disassociation probability is very close to unity in this work. Nitrogen atoms are more readily incorporated into the nanowires than their molecular counterparts and initially incorporated substitutionally at O sites. Given the small nanowire diameters, N atoms will diffuse throughout the entire nanowire volume, whereas gaseous nitrogen adsorbs onto the nanowire surface and then desorbs from the sample by prolonged plasma annealing [13]. As seen in Fig. 3, prolonged plasma annealing did not lead to a significant increase in the overall nitrogen content but instead changed the relative concentrations of nitrogen species. $N_2$ is known to have a lower formation energy than $N_O$ [13], and given the ballistic transport of atoms in nanowires at elevated temperatures, the pairing of N atoms is expected. First principles calculations by Wang and Zunger [32] indicated that $N_O$ is metastable due to an excess energy in the N bonds to the surrounding Zn atoms, leading to the migration of N from this defect at elevated temperatures and the attraction of a second N atom. The observed plasma time-dependent behaviours of the $N_O$ and loosely bound $N_2$ species in this work are consistent with this theoretical prediction.

We propose here that the $N_2$ formation rate is controlled by a vacancy-assisted migration mechanism, provided that the $N_O$ source is not severely depleted. When atomic nitrogen is used as a doping source, the process of forming $N_2$ will be mediated by diffusion-rate kinetics, because two separate N atoms must migrate to the same site. The probability of the $N_2$ pair formation will be proportional to the square of the nitrogen concentration present and thermal energy during annealing:

$$P[\text{NN}] \propto e^{-\frac{E_a}{k_b T}} [\text{N}]^2 \qquad (1)$$

where $E_a$ is the activation energy for the migration of N and [N] denotes the total nitrogen concentration. A related equation similar to Eq. (1) was originally developed for the determination of nitrogen pairs in GaP [33]. Assuming now the same X-ray absorption probability of a N atom for the nitrogen species, the variations in their relative concentrations yield an activation energy $E_a$ = 360 meV, which compares favourably with the binding energy of the $(N_O\text{-}V_O)^+$ complex (310 meV) [10]. Thus, it is highly likely that the formation of $N_2$ is primarily controlled by the migration of $N_O$. Possible differences in photoabsorption cross-sections and the difficulty associated with the measurements of diluted samples are estimated to lead to an uncertainty in $E_a$ of at least 20%. The observed time-dependent behaviour of nitrogen can thus be summarized as follows: nitrogen first substitutionally occupies O sites; however, a N atom paired with another N at Zn sites is thermodynamically more favourable. The dominance of $N_O$ over other occupying sites in the initial stage is due to the fact that $V_O$ is most abundant in the nanowires [15]. The $N_2$ formation process is diffusion rate limited being controlled mainly by kinetics constraints, similar to the formation of N-$V$ defects in nanodiamonds [34]. The transient behaviour of $N_O$ and $(N_2)_{Zn}$ may account for inconsistencies in many reports on N-doped ZnO.

**IV. CONCLUSIONS**

Our data provide new insights into the chemical state of the shallow acceptor in N-doped ZnO. We observed a direct correlation between the intensity of a DAP centered at 3.232 eV and the concentration of $N_2$ molecular species that weakly bond to the ZnO lattice. Based on our experimental evidence, we suggest that the DAP in ZnO is due to recombination involving shallow acceptors that are caused by $(N_2)_{Zn}$, as proposed by Lambrecht and Boonchun [13], and shallow donors. The formation of $N_2$ molecule is found to be controlled by the oxygen vacancy assisted diffusion mechanism. The existence of $N_2$ as a shallow acceptor state has important implications on the growth strategies for *p*-type ZnO.

**ACKNOWLEDGMENTS**

This research was partly undertaken on the Soft X-ray Spectroscopy beamline at the Australian Synchrotron, Victoria, Australia. We wish to acknowledge the excellent technical assistance of Mr Geoffrey McCredie, who constructed the nanowire growth furnace. We thank Prof W. R. L. Lambrecht for stimulating discussion on nitrogen in ZnO.

Figures

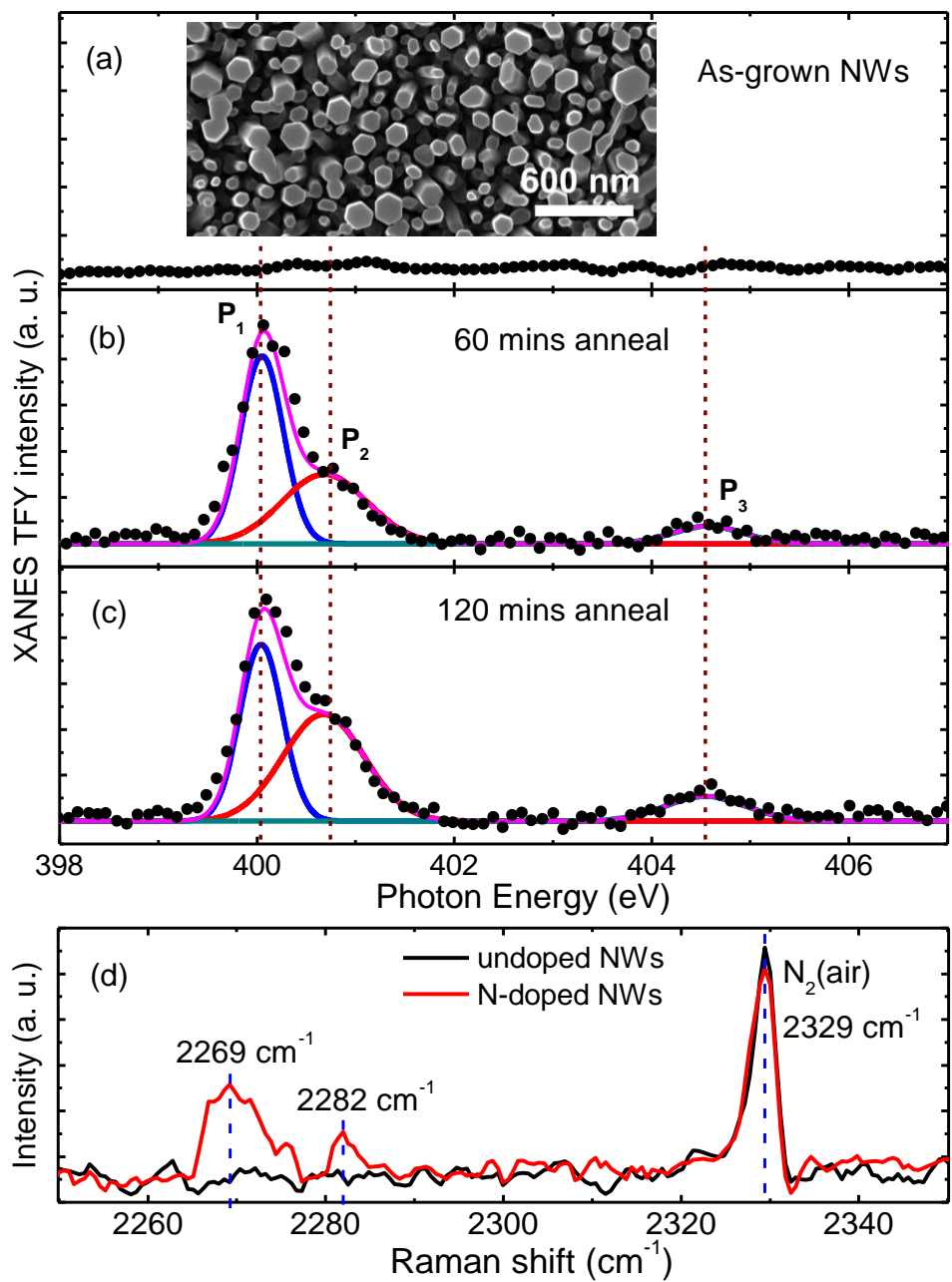

FIG. 1. (a-c) XANES spectra of the as-grown and N-doped ZnO nanowires. The spectra of the N-doped samples could be fitted to three Lorentzian functions (labelled $P_1$, $P_2$ and $P_3$), which are associated with three N-related defects $N_O$, loosely bound $N_2$ and $N_{Zn}$, respectively. Inset shows a typical SEM image of hexagonal-shaped ZnO nanowires used in this study. (d) Raman spectra of undoped and N-doped nanowires, showing the emergence of two high-energy modes at 2269 and 2282 cm$^{-1}$. These peaks, close to the vibrational energy of $N_2$ in air, are assigned to lattice-bound $N_2$ in ZnO.

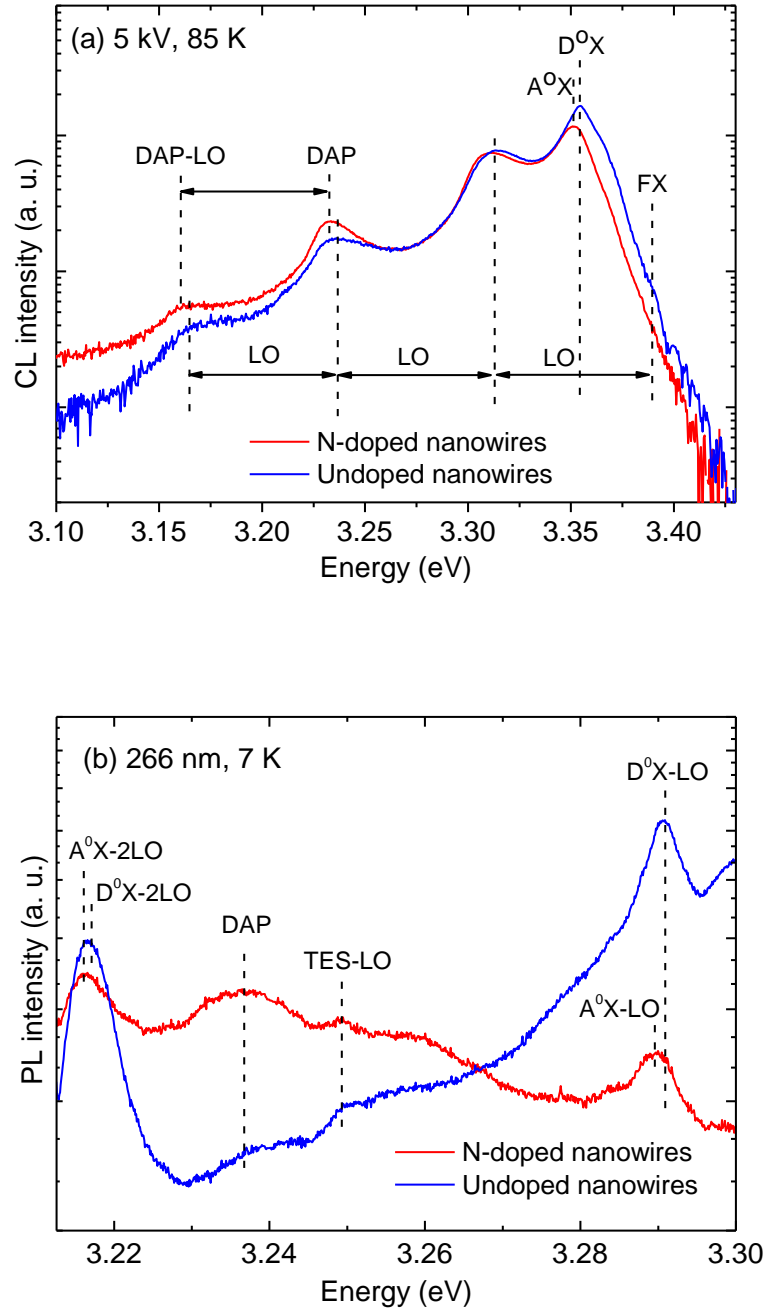

FIG. 2. (a) CL spectra of individual ZnO nanowires measured before and after doping with nitrogen by plasma annealing. The spectra are similar, except the notable enhancement at 3.232 eV corresponding to N-related DAP emission. (b) High resolution PL spectra of nanowire ensembles at 7 K. The spectra are dominated by bound-exciton emissions with a resolvable DAP emission at 3.236 eV in N-doped nanowires.

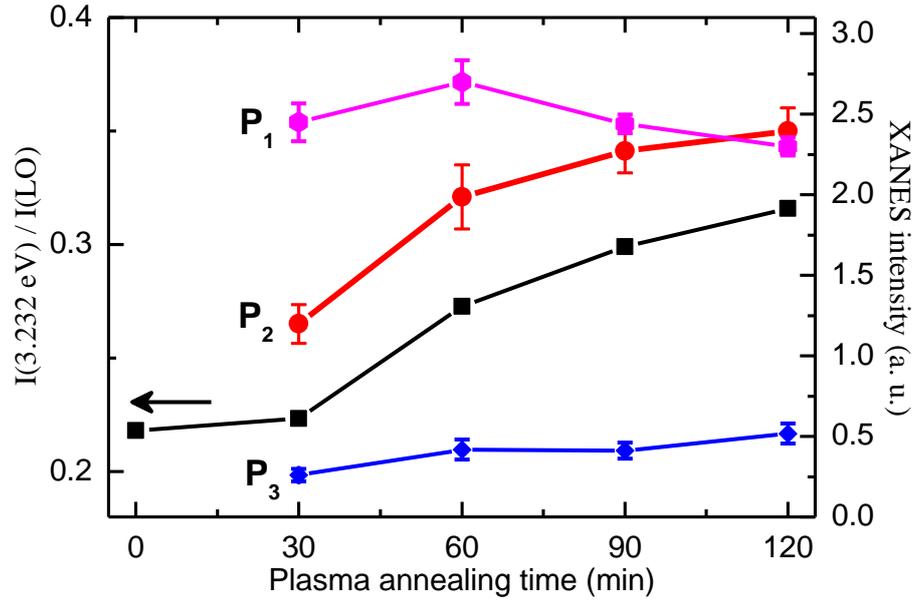

FIG. 3. Dependence of CL peak intensity ratio I(3.232 eV)/I(LO) intensity and XANES components $P_1$, $P_2$ and $P_3$ on plasma annealing time for ZnO nanowires. The direct correlation between the I(3.232 eV) emission and the $P_2$ intensity indicates the involvement of $N_2$ in the DAP emission.

**Appendix A: Raman spectra of N-doped ZnO nanowires**

Raman spectroscopy was employed to investigate the structure of nanowires following nitrogen incorporation. Fig. 4 presents the spectra of undoped and 120 min N-doped ZnO nanowires that were grown on sapphire substrates. The spectra exhibit typical ZnO lattice modes at 438 cm$^{-1}$ ($E_2^{high}$) and 331 cm$^{-1}$ ($E_2^{high} - E_2^{low}$). The strain-sensitive Raman $E_2^{high}$ mode is located at 437.4 ± 0.2 cm$^{-1}$ for both undoped and N-doped nanowires, indicating a minimal compressive strain in the nanowires [35]. No peak shift or broadening of the modes is observed in N-doped nanowires, indicating the structural quality of the nanowires is not reduced by nitrogen doping.

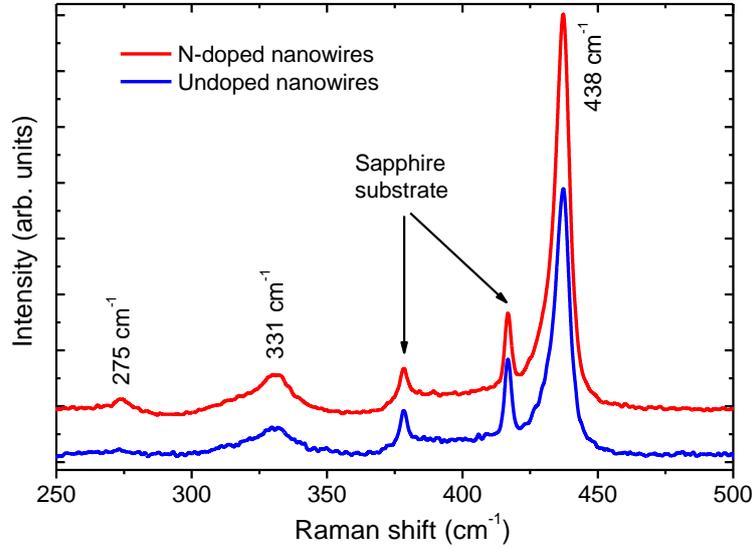

FIG. 4: Raman spectra of undoped and N-doped ZnO nanowires, showing two dominant modes of wurtzite ZnO at 438 cm$^{-1}$ ($E_2^{high}$) and 331 cm$^{-1}$ ($E_2^{high} - E_2^{low}$). The additional 275 cm$^{-1}$ peak in the N-doped spectrum has been attributed to the local vibration of small Zn clusters in the region where oxygen atoms are replaced by nitrogen [36].

**Appendix B: Identification of N-related donor-acceptor pair (DAP) recombination**

Fig. 5 displays representative CL spectra of as-grown and N-doped nanowires at various temperatures between 15 and 140 K, together with the assignments of the spectral features. The phonon replica peaks (FX-LO and FX-2LO) are separated by the longitudinal optical phonon $E_{LO}$ = 72 meV, while at 15 K the energy separation between FX-LO and DAP in N-doped nanowires is 77 meV. The transition energies of FX-LO, FX-2LO and DAP obtained from the CL spectra are plotted in Fig. 6 as a function of temperature. We find the temperature dependencies of FX-LO and FX-2LO are adequately described by the standard equations [26], which are displayed beside each data set in Fig. 6. Due to the thermal energy ½$k_B$T contribution, the FX-2LO and N-related DAP occur essentially at the same energy above 140 K, but gradually separate with decreasing temperature.

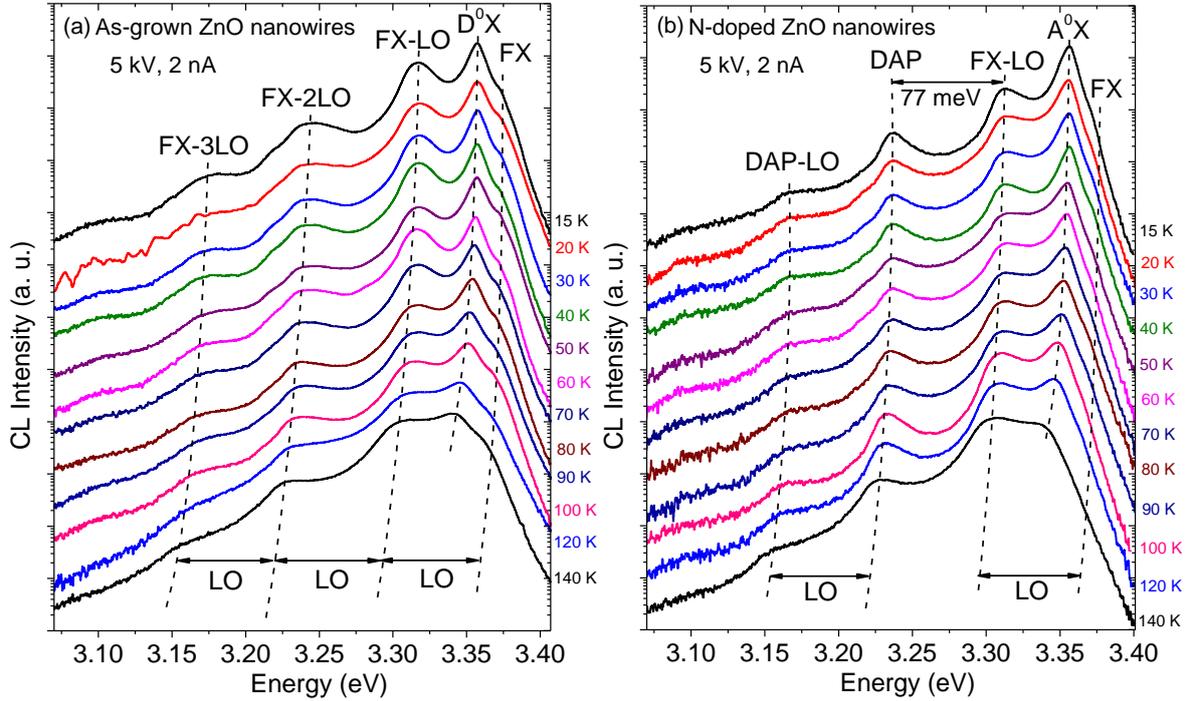

FIG 5. Temperature-resolved CL spectra of (a) as-grown, annealed and (b) N-doped ZnO nanowires ($E_B$ = 5 keV, $I_B$ = 2 nA). The CL intensity is plotted in a logarithmic scale and the spectra are vertically shifted for clarity. The dashed curves indicate the peak positions of $D^0X$, $A^0X$, phonon replicas of FX and DAP. The separation of N-related DAP and FX-LO is 77 meV, significantly greater than the energy of longitudinal optical phonons (72 meV).

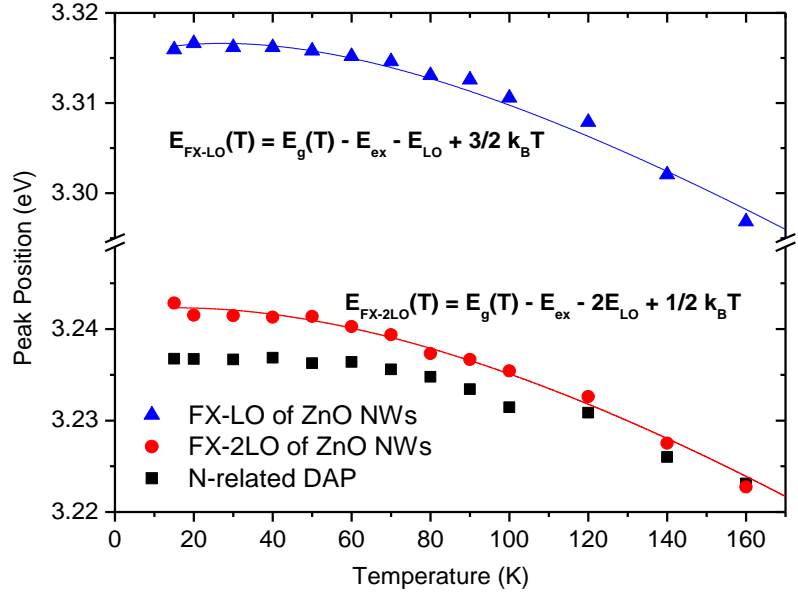

FIG. 6. Temperature dependences of FX-LO, FX-2LO and N-doped DAP transitions energies obtained from the CL spectra. The experimental data are presented by the symbols, whereas the solid lines are simulated curves based on the equations shown in the figure. The fitting parameters used were $E_g(0) = 3.438$ eV, $\alpha = 4.6 \times 10^{-4}$ eV/K, $\beta = 230$ K. The binding energy of free excitons was assumed to be independent of temperature for the temperature range above ($E_{LO} = 72$ meV).

A blue shift of DAP emission peak with increasing excitation density is a common feature of DAP transitions. This shift originates from a redistribution from distant donor-acceptor pairs to closer pairs with stronger Coulomb interaction. The DAP peak positions of the N-doped nanowires obtained from power density-resolved CL spectroscopy at 15 K are presented in Fig. 7 as a function of $e$-beam current. The DAP transition exhibits only a slight blue shift as the current is increased from 0.5 nA to 20 nA. However, a large blue shift is not expected for N-doped ZnO since only a small number of close donor-acceptor pairs exist due to the low concentration of nitrogen acceptors.

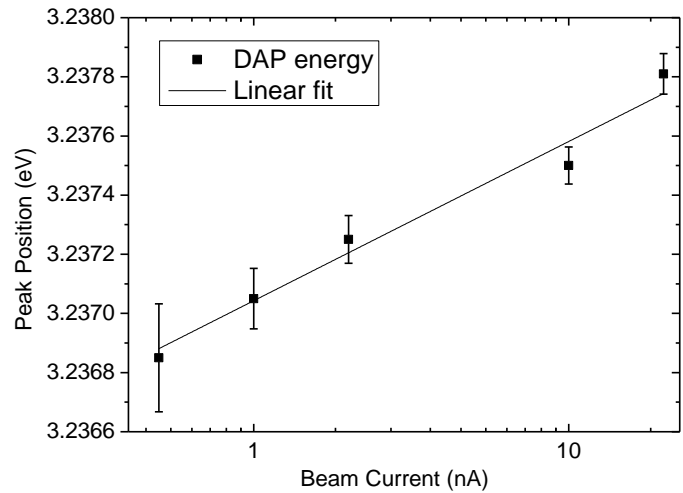

FIG. 7. Excitation density dependence of N-related DAP transition energy, showing a small blue shift with increasing *e*-beam current.